\documentclass[10pt,twocolumn,floatfix]{revtex4-1}
\usepackage{graphicx}
\usepackage{epstopdf}
\usepackage{amssymb}
\usepackage{mathrsfs}
\usepackage{amsmath}
\usepackage{color}
\usepackage{latexsym}
\usepackage{amsfonts}
\usepackage{hyperref}
\usepackage{subfigure}
\usepackage{wasysym}
\usepackage{dcolumn}
\usepackage{bm}
\usepackage{hyperref}

\begin{document}

  
\title{How the stability of a folded protein depends on interfacial
  water properties and residue-residue interactions}

\author{Valentino Bianco$^{1}$}
\email{valentino.bianco@univie.ac.at, gfranzese@ub.edu}


\author{Neus Pag\`es Gelabert$^2$}

\author{Ivan Coluzza$^1$}

\author{Giancarlo~Franzese$^{2,3}$}
\email{valentino.bianco@univie.ac.at, gfranzese@ub.edu}



\affiliation{$^1$
  Universit\"at Wien, Sensengasse 8/10, 1090 Vienna, Austria}

\affiliation{$^2$
Secci\'o de
    F\'isica Estad\'istica i Interdisciplin\`aria--Departament de
    F\'isica de la Mat\`eria Condensada, Facultat de F\'isica
    Universitat de Barcelona,  
    Mart\'{\i} i Franqu\`es 1, 08028, Barcelona, Spain}

\affiliation{$^3$
  Institute of Nanoscience and Nanotechnology (IN2UB), Universitat de Barcelona, Av. Joan XXIII S/N, 08028 Barcelona,
    Spain}

\date{\today}

\definecolor{corr14okt}{rgb}{0,0,1} 
\definecolor{moved}{rgb}{0,0,0}

\begin{abstract}
  Proteins work only if folded in their native state, but changes in
  temperature $T$ and pressure $P$ induce their unfolding.  Therefore
  for each protein there is a  stability region (SR) in the $T$--$P$
  thermodynamic plane outside which the biomolecule is denaturated.
  It is known that the extension and shape of the SR
  depend on i) the specific protein residue-residue interactions in the
  native state of  the amino acids sequence and ii) the water
  properties at the hydration interface. Here we analyze by Monte
  Carlo simulations of different coarse-grained protein models in explicit water
  how changes in i) and ii) affect the  SR. We show that the solvent
  properties ii) are essential to rationalize the SR shape at low $T$
  and high $P$ and that our finding are robust with respect
  to parameter changes and with respect to different protein
  models. These results can help in developing new strategies for
  the design of novel synthetic biopolymers. 
\end{abstract}  


\maketitle


\section{Introduction}
The capability of the single components to independently organize in
pattern and structures without an external action fulfils a crucial
role in the supramulecular organization and assembling of the
biological matter \cite{Lehn2002, Whitesides2002a}. To cite some
examples, self-assembly is observed in bio--molecules \cite{Yin2008a},
in DNA and chromosomes \cite{Carroll2006, BiancoBPJ2012, Douglas2009a,
  DeMichele2012}, in lipid membranes \cite{Antonietti2003, Calero2016}, in the
cytoskeleton \cite{Nedelec2003}, in cells and tissues
\cite{Gerecht-Nir2004, Jakab2007}, in virus and bacteria
\cite{Rong2011, Sakimoto2014}, and in proteins
\cite{VanderLinden2007, Vilanova2016}. In particular, the protein folding represents
one of the most challenging and elusive biochemical processes where a
chain of amino acids organizes itself into a unique native and folded
structure \cite{Dobson2003, Finkelstein2004}. 
The protein folding is a spontaneous process driven by intra-molecular
(residue-residue) van der
Walls interactions and hydrogen bonds which overcome the
conformational entropy. It depends also on the presence of  co-factors
as the chaperones \cite{Coluzza2006a} and, in particular, the properties of
the solvent, i.e.  water \cite{BiancoPRL2015}, and the co-solutes
\cite{Ghosh2017}
that regulate the pH
level and the salt concentration, for example.

Although water has no influence on the primary structure (the protein
sequence), it affects the protein in all the other level of organization
\cite{LevyPNAS2004, Levy2006, Raschke2006}. Indeed,  
i) water forms H-bonds with the  polar/charged residues of the side
chains, influencing the adoption of secondary structures like alpha
helices or beta sheets which expose the most hydrophilic residues to
water; ii) the hydrophobic effect drives the collapse of the protein
core and stabilizes the tertiary protein structure; iii) water induces
the aggregation of proteins since they usually present hydrophobic
regions on their surface (quaternary structure).

Experiments have clearly documented that 
proteins maintain their native structure in a limited range of
temperatures $T$ and pressures $P$ \cite{Zipp1973, privalov,
  HummerPNAS1998, MeersmanHighPressRes2000, Lassalle2000, Smeller2002,
  Herberhold2002, Lesch2002, RavindraChemPhysChem2003,
  MeersmanChemSocRev2006, PastoreJACS2007, WiedersichPNAS2008,
  Maeno2009, Somkuti2013, Somkuti2013a, NucciPNAS2014} showing an
elliptic-like stability region (SR) in the $T$--$P$ plane, as
accounted by a Hawley's theory \cite{hawley}. Outside its SR
a protein unfolds, with a
consequent loss of its tertiary structure and functionality.

At high $T$ the protein unfolding is due to the thermal fluctuations
which disrupt the protein structure. Open protein conformations
increases the entropy $S$ minimizing the global Gibbs free energy
$G\equiv H-TS$, where $H$ is the total enthalpy. 
Upon cooling, if the nucleation of water is avoided, some proteins
cold--denaturate \cite{privalov, Griko1988, RavindraChemPhysChem2003,
  goossens, nash, nash2, MeersmanHighPressRes2000,
  PastoreJACS2007}. Usually such a phenomenon is observed below the
melting line of water, although in some cases cold denaturation occurs
above the $0^\circ$C, as in the case of the yeast frataxin
\cite{PastoreJACS2007}.  
Protein denaturation is observed, or predicted, also upon
pressurization \cite{Zipp1973, HummerPNAS1998, PaschekPRL2004,
  MeersmanChemSocRev2006, NucciPNAS2014}. A possible explanation of
the high-$P$ unfolding is the loss of  internal cavities, sometimes presents in
the folded states of proteins \cite{RochePNAS2012}. Denaturation at
negative $P$ has been experimentally observed \cite{Larios2010} and
simulated recently \cite{Larios2010, HatchJPCB2014, BiancoPRL2015}.
Pressure denaturation is usually observed for 100 MPa $\lesssim P \lesssim$ 600 MPa, and rarely at higher $P$ unless the tertiary structure is engineered with stronger covalent bonds \cite{Lesch2002}.
Cold- and $P$-denaturation of proteins have been related to the equilibrium properties of
the hydration water \cite{delosriosPRE2000, marquesPRL2003, PatelBPJ2007, Athawale2007, NettelsPNAS2009, Best2010, Jamadagni2010, Badasyan2011, matysiakJPCB2012, BiancoJBioPhys2012, BiancoPRL2015}.
However, the interpretations of the mechanism is still largely debated
\cite{PaschekPRL2004, Paschek2005, Sumi2011, Coluzza2011, Dias2012,
  DasJPCB2012, SarmaCP2012, FranzeseFood2013, Abeln2014, Yang2014,
  RochePNAS2012, Nisius2012, VanDijk2016a}.

Here we investigate by Monte Carlo simulations of different
coarse-grained protein models in explicit water how the SR is affected
by changes in i) the specific protein residue-residue interactions in the native
state of the amino acids sequence and ii) the solvent properties at the
hydration interface, focusing on water energy and density fluctuations.
In particular, after introducing the model and the numerical method in
Section II, we study in a broad range of $T$ and $P$ how
the conformational space of proteins depends on the model's parameters
for the hydration water in Section III.A 
and how it depends on the residue-residue interactions in Section
III.B. Next, we discuss the possible relevance of our results in the framework
of protein design in Section IV and, finally, 
we present our concluding remarks
in Section V.


\section{Models and Methods}
 
The extensive exploration with  atomistic models of  protein conformations in
explicit solvent at different
thermodynamic conditions, including extreme low $T$ and high $P$,
is a very demanding analysis. To overcome this limitation, we adopt a
coarse-grain model 
for protein-water interaction based on
A) the {\it
  many-body} water model \cite{StokelyPNAS2010, strekalovaPRL2011, oriol,
  FranzeseFood2011, MazzaPNAS2011, BiancoJBioPhys2012,
  FranzeseFood2013, BiancoProceedings2013, delosSantos2011,
  BiancoSR2014, BiancoPRL2015, Enrique-preprint-2016},
combined with
B) a lattice representation of the
  protein.

The many-body water model has been proven to
reproduce--in at least qualitative way--the thermodynamic
\cite{StokelyPNAS2010, Enrique-preprint-2016} and dynamic \cite{delosSantos2011} behavior
of water, the properties of water in confinement
\cite{strekalovaPRL2011, oriol, BiancoProceedings2013, BiancoSR2014} and at
the inorganic interfaces \cite{FranzeseFood2013}. Its recent combination with
the lattice representation of the  protein has given a novel
insight into the water-protein interplay \cite{FranzeseFood2011,
  MazzaPNAS2011, BiancoJBioPhys2012, BiancoPRL2015}. 

As we will describe later, for the protein we consider a 
model that, in its general formulation as polar protein,
follows the so-called ``Go-models'',
a common approach in protein folding.
In their seminal
paper Go
and Taketomi~\cite{GO1978} employed non-transferable potentials
tailored to the native structure. The interactions were designed to
have a sharp minimum only at the native residue-residue distance,
guaranteeing that the energy minimum is reached only by the native
structure.  The Go-proteins thus successfully fold, and have a smooth
free-energy landscape with a single global minimum in the native
structure~\cite{Cheung2004}. Hence, Go-models are equivalent to having
an infinite variety of pair interactions among the residues (alphabet
$\mathscr A$), such that each amino acid interacts selectively with a
subset of residues defined by the distances in the native
configuration. If the size of the alphabet is reduced, the
construction of folding proteins requires  an optimization step of
amino acid sequence along the chain~\cite{Shakhnovich1993c,
  Coluzza2011, Coluzza2014}; for this reason these methods are often
referred as ``protein design''. Comparing designed proteins with
Go-proteins, Coluzza recently shown that, close to the folded state, Go
and designed proteins behaves in a very similar manner
\cite{Coluzza2015}. Since we are interested in measuring the stability
regions defined by the environmental condition at which the trial
protein is at least 90\% folded, Go-models are an appropriate protein
representation, and, at this stage, we do not require to perform the
laborious work of protein design to obtain general results. We will
discuss later the possibility to extend our model to the case of a
limited alphabet $\mathscr A$ of residues (20 amino acids).

\subsection{The bulk many-body water model.} 

We consider the coarse-grain many-body bulk water at constant $P$,
constant $T$ and constant number $N^{\rm (b)}$ of  water molecules, while the
total volume $V^{\rm (b)}$ occupied by water is a function of $P$ and
$T$. Because in the following we will consider the model with 
water at the hydration protein
interface and (bulk) water away from the interface, for sake of
clarity here we introduce the notation with a superscript ${\rm (b)}$ for
quantities that refer to the bulk.

We replace the  coordinates and orientations of the water molecules 
by a continuous density field and discrete
bonding variables, respectively. The density field is defined based on
a partition of the available volume  $V^{\rm (b)}$ into a fixed
number $N_0=N^{\rm (b)}$ of cells, each with volume $v^{\rm (b)}\equiv V^{\rm (b)}/N^{\rm (b)}\geq v_0$, where
$v_0\equiv r_0^3$ is the water excluded volume with
$r_0\equiv 2.9 $\AA{}  (water van der Waals 
diameter).  For sake of simplicity we assume
that, when the water molecules are not forming hydrogen bonds (HBs), the (dimensionless)
density is homogeneous in each cell and equal to $\rho^{\rm (b)}\equiv v_0/v^{\rm (b)}$.
As we will discuss later, the density is, instead, locally inhomogeneous
when  water molecules form HBs. Specifically,
the density depends on the number of HBs,
therefore $\rho^{\rm (b)}$
only represents the average bulk density.

The Hamiltonian of the bulk water is
\begin{equation}
\mathscr{H}^{\rm (b)}\equiv \sum_{ij} U(r_{ij}) -J N_{\rm HB}^{\rm (b)}-J_\sigma N_{\rm  coop}^{\rm (b)}.
\label{bulk}
\end{equation}
The first term represents the isotropic part of the water-water
interaction and accounts for the van der Waals interaction
\cite{Morawietz2016}. It is modeled with a Lennad-Jones potential  
\begin{equation}
  \sum_{ij} U(r_{ij})\equiv 4\epsilon\sum_{ij}\left[\left(\frac{r_0}{r_{ij}}\right)^{12}-\left(\frac{r_0}{r_{ij}}^6\right)\right]
\end{equation}
where $\epsilon \equiv 5.8$ kJ/mol and
the sum runs over all the water molecules $i$ and $j$ at O--O
distance $r_{ij}$ calculated as the distance between the centers of
the two cells $i$ and $j$ where the molecules belong.
We assume a hard-core exclusion 
$U(r)\equiv \infty$ for $r<r_0$ and a cutoff 
for $r>r_c \equiv 6 r_0$.

The second term in Eq. (\ref{bulk}) represents the directional (covalent) component of the HB, where 
\begin{equation}
 N_{\rm HB}^{\rm (b)}\equiv\sum_{\langle ij \rangle}n_i
 n_j\delta_{\sigma_{ij},\sigma_{ji}}
 \label{NHB}
\end{equation}
is the number of bulk HBs and the sum runs over neighbor cells
occupied by water molecules. Here we introduce the label $n_{i}=1$ if
the cell $i$ has a water density $\rho^{\rm (b)}>0.5$
and $n_{i}=0$ otherwise. In the homogeneous bulk this condition
guarantees that two water molecules can form a HB only if their relative 
distance is $r<2^{1/3}r_0\equiv 3.66$ \AA, corresponding to the 
range of a water's first coordination shell  as
determined from the O-O radial distribution function from 220 to 673 K
and at pressures up to 400 MPa  \cite{Soper:2000jo}.

The variable $\sigma_{ij}=1, \dots, q$  in Eq. (\ref{NHB})  is the bonding index of the water
molecule in cell $i$ with respect to the neighbor molecule in cell
$j$ and $\delta_{ab}=1$ if $a=b$, or 0 otherwise, is a Kronecker delta function.
Each water molecule has as many bonding variables as neighbor cells,
but  can form only up to four HBs.
Therefore, if the molecule has more than four neighbors, e.g., in a
cubic lattice partition of $V^{\rm (b)}$, an additional condition must
be applied to limit to four the HBs participated by each molecule.

The parameter $q$ in the definition of 
$\sigma_{ij}$ is determined by the entropy decrease associated to the
formation of each HB. 
Each HB is unbroken if the hydrogen atom H is in a
range of $[-30^\circ, 30^\circ]$ with respect to the O--O axes
\cite{Teixeira1990}.
Hence,
only $1/6$ of the entire range of values $[0,360^\circ]$ for  the
${\widehat{\rm OOH}}$ angle is associated to a bonded state.
Therefore, in the zero-order approximation of considering each HB
independent, a molecules that has $4-n$ HBs, with $n=1, \dots 4$, has
an orientational entropy that is $S^{\rm o}_n/k_B\equiv n \ln 6$ above
that of a
fully bonded molecule with $S^{\rm o}_0/k_B\equiv 0$, where $k_B$ is
the Boltzmann constant. As a consequence, the choice $q=6$   accounts
correctly for the entropy variation due to  HB formation and breaking
given the standard definition of HB.

The third term in Eq.~(\ref{bulk})
is associated to the  cooperativity of the HBs due to the quantum many-body interactions
\cite{HernandezJACS2005, StokelyPNAS2010}. Indeed, the formation of a new HB affects
the electron probability distribution around the molecule favoring the
formation of the following HB in a local tetrahedral structure
\cite{SoperPRL2000}.
We assume that the energy gain due to this effect 
is proportional to the number of cooperative HBs  in the system
\begin{equation}
 N_{\rm coop}\equiv \sum_i
 n_i\sum_{(l,k)_i}\delta_{\sigma_{ik},\sigma_{il}},
 \label{NCOOP}
\end{equation}
where $n_i$  assures that we include this term only for liquid water.
With this definition and with the choice $J_\sigma/4\epsilon\ll J$ the term
mimics a many-body interactions among the HBs participated by the same
molecule. Indeed, the condition $J_\sigma/4\epsilon\ll J$ guarantees
that the interaction takes place only when the water molecule $i$ is
forming several HBs. The inner sum is over  
$(l,k)_i$, indicating each of the six different pairs of the four
indices $\sigma_{ij}$ of the molecule $i$.

The formation of HBs leads to an open network of molecules, giving
rise to a lower density state. We include this effect into the model
assuming that for each HB  the volume $V^{\rm (b)}$ increases of
$v_{\rm  HB}^{\rm (b)}/v_0=0.5$.
This value is the average volume increase between 
high-density ices VI and VIII and low-density (tetrahedral) ice Ih.
As a consequence, the average bulk density is 
\begin{equation}
 \rho^{\rm (b)}\equiv \frac{Nv_0}{V^{\rm (b)}+N_{\rm HB}^{\rm (b)}v_{\rm HB}^{\rm (b)}}.
\end{equation}
We
assume that the HBs do not affect the distance $r$ between first
neighbour molecules, consistent with experiments
\cite{SoperPRL2000}. Hence, the water-water distances $r$ 
is calculated only from $V^{\rm (b)}$.

As discussed in Ref.~\cite{BiancoPRL2015}
a good choice for the parameters that accounts for the ions in a protein 
solution 
is $\epsilon =5.8$ kJ/mol,
$J/4\epsilon=0.3$ and $J_\sigma/4\epsilon = 0.05$ that give an average
HB energy $\sim 20$ kJ/mol. In the following 
 we consider two protein models, a simpler one used to understand
the molecular mechanisms through which water contributes to the
unfolding, and a more detailed model which includes the effect of
polarization. For sake of simplicity, we present here the result for a
system in two dimension. Preliminary results for the model in three
dimensions of both bulk water \cite{Enrique-preprint-2016} and protein folding
 show results that are qualitatively similar to those
presented here.  

\begin{figure}
\begin{center}
 \includegraphics[scale=0.2]{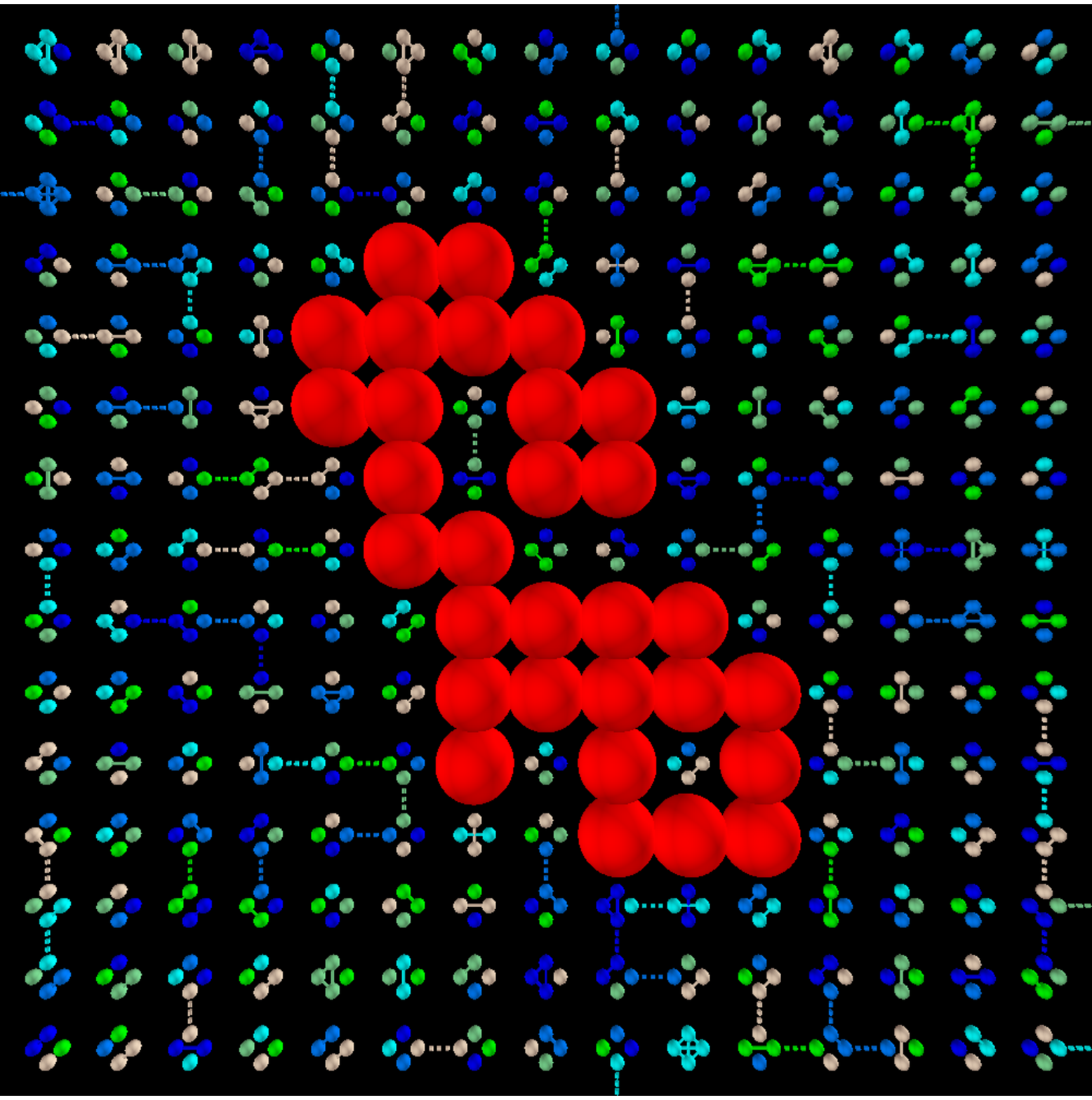}
\end{center}
 \caption{Scheme of the water-protein coarse grain model. The protein
   is represented with red spheres. Each water molecules is
   represented through its 4 bonding indexes $\sigma$, with different
   colours associated to the value $1...q$ assumed by
   $\sigma$. Directional HB are represented with dotted lines joining
   two water molecules. Cooperative bonds are represented with
   continuous lines connecting the $\sigma$ indices inside a
   molecule.} 
 \label{model}
\end{figure}

\subsection{Hydrophobic protein model.} 

The protein is modelled as a
self-avoiding lattice
polymer, embedded into the cell partition of the system. Despite its
simplicity, lattice protein models are still widely used in the
contest of protein folding \cite{lauMacromol1989,
  caldarelliJBioPhys2001, marquesPRL2003, Coluzza2003, PatelBPJ2007,
  matysiakJPCB2012, BiancoPRL2015, VanDijk2016a} because of their
versatility and the possibility to develop coarse-grained theories and simulations
for them. 
Each protein residue
(polymer bead) occupies one cell.
In the
present study, we do not consider the presence of cavities into the
protein structure. 

To simplify the discussion in this first part of the work, we assume
that (i) there is no residue-residue interaction, (ii)  the residue-water
interaction
vanishes, unless otherwise
specified and (iii) all the residues are hydrophobic.
This implies that the protein has multiple ground states, all with the
same maximum number $n_{\rm max}$ of residue-residue contacts.   
As shown by Bianco and Franzese \cite{BiancoPRL2015}, the 
results hold also when the hypothesis (i), (ii) and (iii) are
released, as we will discuss in the following.

Our stating hypothesis is that the protein interface affects the
water-water properties in the hydration shell, here defined as the
layer of first neighbour water molecules in contact with the protein
(Fig. \ref{model}).  
There are many numerical and experimental evidences supporting this
hypothesis.
In particular, it has been shown 
that the water-water HBs in the protein hydration shell are more stable and
more correlated with respect to the bulk HBs \cite{DiasPRL2008,
  PetersenJCP2009, SarupriaPRL2009, TarasevichCollJ2011, DavisNat2012,
  laage2017}.  
We account for this by replacing $J$ of Eq.~(\ref{bulk}) with
$J_{\Phi}>J$ for the water-water HBs at the hydrophobic ($\Phi$) interface. 
Another possibility, discussed later, would be to consider that the
cooperative interaction $J_{\sigma,\Phi}$ at the $\Phi$-interface,
directly related to the tetrahedral order of the water molecules, is
stronger with respect to the bulk. This case would be consistent with
the assumption that water forms ice-like cages around $\Phi$-residues \cite{Grdadolnik2017}.
Both choices, according to Muller discussion \cite{muller1990}, would
ensure the water enthalpy compensation during the cold-denaturation 
\cite{BiancoJBioPhys2012}. 

At the $\Phi$-interface, beside the stronger/stabler water-water HB,
we consider also the larger density fluctuations with respect to the
bulk, as observed in hydrated $\Phi$-solutes
\cite{SarupriaPRL2009, DasJPCB2012}.
As a consequence, at ambient pressure $\Phi$-hydration water is more
compressible than bulk water.

Although it is still matter of debate if, at ambient
conditions, the average density of water at the $\Phi$-interface is
larger or smaller with respect to the average bulk water density
\cite{Lum:1999kx, Schwendel2003, jensenPRL2003, DoshiPNAS2005,
  Godawat2009}, there are evidences showing that such  density
fluctuations reduce upon pressurization \cite{SarupriaPRL2009,
  DasJPCB2012, GhoshJACS2001, DiasJPCB2014}. We include this effect in
the model by assuming that the volume change $v_{\rm HB}^{(\Phi)}$ associated to the HB
formation in the $\Phi$ hydration shell can be expanded 
as  a series function of $P$
\begin{equation}\label{vsurf}
 v_{\rm HB}^{(\Phi)}/v_{\rm HB,0}^{(\Phi)}\equiv 1- k_1P - k_2P^2 -k_3P^3 + O(P^4)
\end{equation} 
where $v_{\rm HB,0}^{(\Phi)}$ is the value of the change when 
$P=0$. Here 
the coefficients $k_1$, $k_2$ and $k_3$ are such that
$\partial v_{\rm HB}^{(\Phi)}/\partial P$ is always negative. 
As first approximation, we study the linear case, with
$k_i=0$ $\forall i>1$.
We discuss later how the protein stability is affected by
considering the quadratic terms in Eq.(\ref{vsurf}).
Our initial choice implies that we can study the system only when
$P<1/k_1$. As we will discuss in the next section, this condition
does not limit the validity of our results.
The total volume $V$ of the system is, therefore, 
\begin{equation}
 V\equiv Nv_0+N_{\rm HB}^{\rm (b)}v_{\rm HB}^{\rm (b)} +N_{\rm HB}^{(\Phi)}v_{\rm HB}^{ (\Phi)}.
\label{Vtot}
\end{equation}
where $N_{\rm HB}^{(\Phi)}$ is the number of HBs in the $\Phi$ shell.

\subsection{ Polar protein model}

In order to account for the effect of the hydrophilic residues on the
water-water hydrogen bonding in the hydration shell, we consider also
the case in which the 
protein is modeled as a heteropolymer composed by hydrophobic ($\Phi$) and
hydrophilic ($\zeta$) residues.  In this case is worth introducing
residue-residues interactions that lead to a specific folded (native) state for
the protein. 

We fix the native state by defining the interaction
matrix $A_{i,j}\equiv \epsilon_{\rm rr}$ if residues $i$ and $j$ are n.n. in
the native state, $0$ otherwise.
To simplify our model we set 
all the residues in contact with
water in the native state as hydrophilic,
and  all those buried into
the protein core as hydrophobic. 
The water interaction with $\Phi$- and  $\zeta$-residues is given by
the parameters $\epsilon_{\rm w,\Phi}$ and $\epsilon_{\rm w,\zeta}$
respectively, where we assume $\epsilon_{\rm w,\Phi}<J$ and
$\epsilon_{\rm w,\zeta}>J$.

The polar $\zeta$ residues interfere with the formation of HB of the
surrounding molecules, disrupting the tetrahedral order and distorting
the HB network.  
Thus we assume that each $\zeta$ residue has a preassigned bonding
state $q^{(\zeta)}=1,...,q$, different and random for each $\zeta$
residue. In this way, a water molecule $i$ can form a HB with a
$\zeta$  residue, located in the direction $j$, only if
$\sigma_{i,j}=q^{(\zeta)}$.

In the polar potein model, the formation of water-water HBs in the hydration shell is described
by the parameters i) $J_\Phi$ and $J_{\sigma,\Phi}$ (directional and
cooperative components of the HB) if both molecules hydrates two
$\Phi$-residues; ii) $J_\zeta$ and $J_{\sigma,\zeta}$ if both
molecules hydrates two $\zeta$-residues; iii) $J_{\Phi,\zeta} \equiv
(J_\Phi+ J_\zeta)/2$ and $J_{\sigma, \Phi,\zeta}\equiv
(J_{\sigma,\Phi} + J_{\sigma,\zeta})/2$ if the two water molecules are
in contact one with a $\Phi$-residue and another with a $\zeta$-residue, forming a
$\Phi$-$\zeta$-interface. Accordingly, the volume associated to the
formation of HB 
in the hydration shell is $v_{\rm HB}^{(\Phi)}$, $v_{\rm
  HB}^{(\zeta)}$ and $v_{\rm HB}^{(\Phi,\zeta)}$. Then, we assume that
$v_{\rm HB}^{(\Phi)}$ changes with $P$ following the
Eq. (\ref{vsurf}). Due to the condition
$\epsilon_{\rm w,\zeta}>J$, we assume that the density
fluctuations near a 
$\zeta$-residue 
are comparable, or smaller, than those in  bulk water, therefore we set $v_{\rm
  HB}^{(\zeta)}=v_{\rm HB}^{(b)}$. Finally, we define $v_{\rm
  HB}^{(\Phi,\zeta)} \equiv (v_{\rm HB}^{(\Phi)} + v_{\rm
  HB}^{(\zeta)})/2$.

\subsection{Simulations' details}
We study proteins of 30 residues with Monte Carlo simulations in the
isobaric-isothermal ensamble, i.e. with constant $P$, constant $T$ and
constant number of particles.  
Along the simulation we calculate the average number of
residue-residue contacts to estimate the protein compactness, sampling
$\sim 10^5$ independent protein conformations for each thermodynamic
state point. For the hydrophobic protein model, we assume that the
protein is folded if the average number of residue-residue contacts is
$n_{\rm rr}\geq 50\% ~n_{\rm  max}$, while for the polar protein
model, having a unique folded state, we fix the threshold at $n_{\rm
  rr}\geq 90\% ~n_{\rm  max}$.

For sake of simplicity, we consider our model in two
dimensions. Although this geometry could appear as not relevant for
experimental cases, our preliminary results for the three dimensional
system show no qualitative difference with the case presented here. We
understand this finding as a consequence of the peculiar property of
bulk water of having, on average, not more than four neighbors. This
coordination number is preserved if we consider  a square partition of a two
dimensional system. Differences between the two dimensional and the
three dimensional models could arise from the larger entropy in higher
dimensions for the protein, however our preliminary results in 3D show that they can
be accounted for by tuning the model parameters.

\begin{figure}
\begin{center}
 \includegraphics[scale=0.4,bb=4 10 598 446,clip=true]{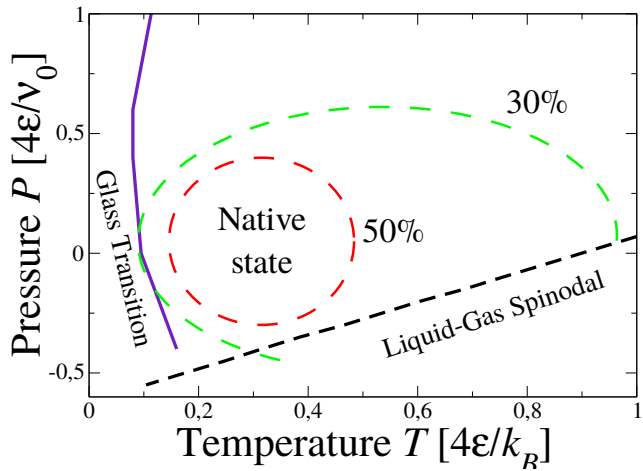}
 \end{center}
\caption{Stability region for a coarse-grained proteins made of 30
   hydrophobic residues. The dotted green line delimits the region within which the
   protein makes at least 30\% of its maximum number of contact
   points, i.e.,  $n_{\rm rr}/n_{\rm  max} \geq 0.3$. Inside this
   region, the  dotted red line delimits the set of states for which $n_{\rm
     rr}/n_{\rm  max} \geq 0.5$, that by definition correspond to the
   native state of the folded protein. The lower straight 
   (black) dotted line represents the limit of stability (spinodal) of the liquid water with
   respect to the gas. The left-most (violet) solid line  marks the
   limit below which water forms a glass state.
 Adapted from Ref. \cite{BiancoPRL2015}.}
 \label{SR0}
\end{figure}

\section{Results and discussion}

\subsection{Results for the hydrophobic protein model}

Bianco and Franzese
show \cite{BiancoPRL2015} that
the hydrophobic protein model, with 
 parameters $k_1= v_0/4\epsilon$ (and $k_2=k_3=0$),
 $v_{\rm HB,0}^{(\Phi)}/v_0=v_{\rm HB}^{\rm (b)}/v_0=0.5$ and
 $J_{\Phi}/4\epsilon=0.55$,
$J_{\sigma,\Phi}=J_\sigma$,
 has a SR that is  elliptic in
 the $T$--$P$ plane. This finding is consistent with  the predictions
 of the Hawley
 theory \cite{hawley,Smeller2002} accounting 
 for the thermal, cold and pressure denaturation
 (Fig. \ref{SR0}).

 They find that at high $T$ the large entropy associated to open
 protein conformations keeps the protein unfolded.
 By isobaric decrease of $T$, the energy
 cost of an extended water-protein interface can no longer be balanced
 by the entropy gain of the unfolded protein, and 
 the protein  folds to  
 minimizes the number of hydrated $\Phi$-residues, as expected.

 By further decreasing of $T$ at constant $P$, the number of
 water-water HBs increases both in bulk and at the protein
 interface. At low-enough $T$, the larger stability, i.e., larger energy gain,
 of the HBs at the $\Phi$-interface drives the cold denaturation of the protein.

 Upon isothermal increase of $P$, the enthalpy of the system increases
 for the increasing $PV$ term. Therefore, a mechanisms that reduces
 $V$ would reduce the total enthalpy. Here the mechanism is provided
 by the water compressibility that is larger at the $\Phi$-interface
 than in bulk. Therefore, the larger water density at the protein
 interface drives the unfolding, which leads to a larger
 $\Phi$-interface and enthalpy gain.
 
Finally, when the system is under tension, i.e., at $P<0$, the total
enthalpy is minimized when $V$ in Eq.(\ref{Vtot}) is 
maximized. However, the increase of average separation between water
molecules breaks the HBs. In particular, bulk HBs break more than those at
the $\Phi$-interface because the first  are weaker than the
latter. Hence, $N_{\rm HB}^{\rm (b)}$ vanishes when
$N_{\rm  HB}^{(\Phi)}>0$. As a consequence, the maximization of $V$ is
achieved by maximizing $N_{\rm  HB}^{(\Phi)}$, i.e., by exposing the
maximum number of $\Phi$-residues, leading to the protein denaturation
under tension.

Once it is clear that the model can reproduce the protein SR,
allowing us to understand the driving mechanism for the denaturation at
different thermodynamic conditions, it is insightful to study how the
SR depends on the model parameters. Therefore, in the following
of this work we show our new calculations about the effect of varying one
by one the model parameters.

\begin{figure}
\begin{center}
 \leftskip 0.4cm {\bf (a)}\\
\includegraphics[scale=0.39,bb=0 0 668 450,clip=true]{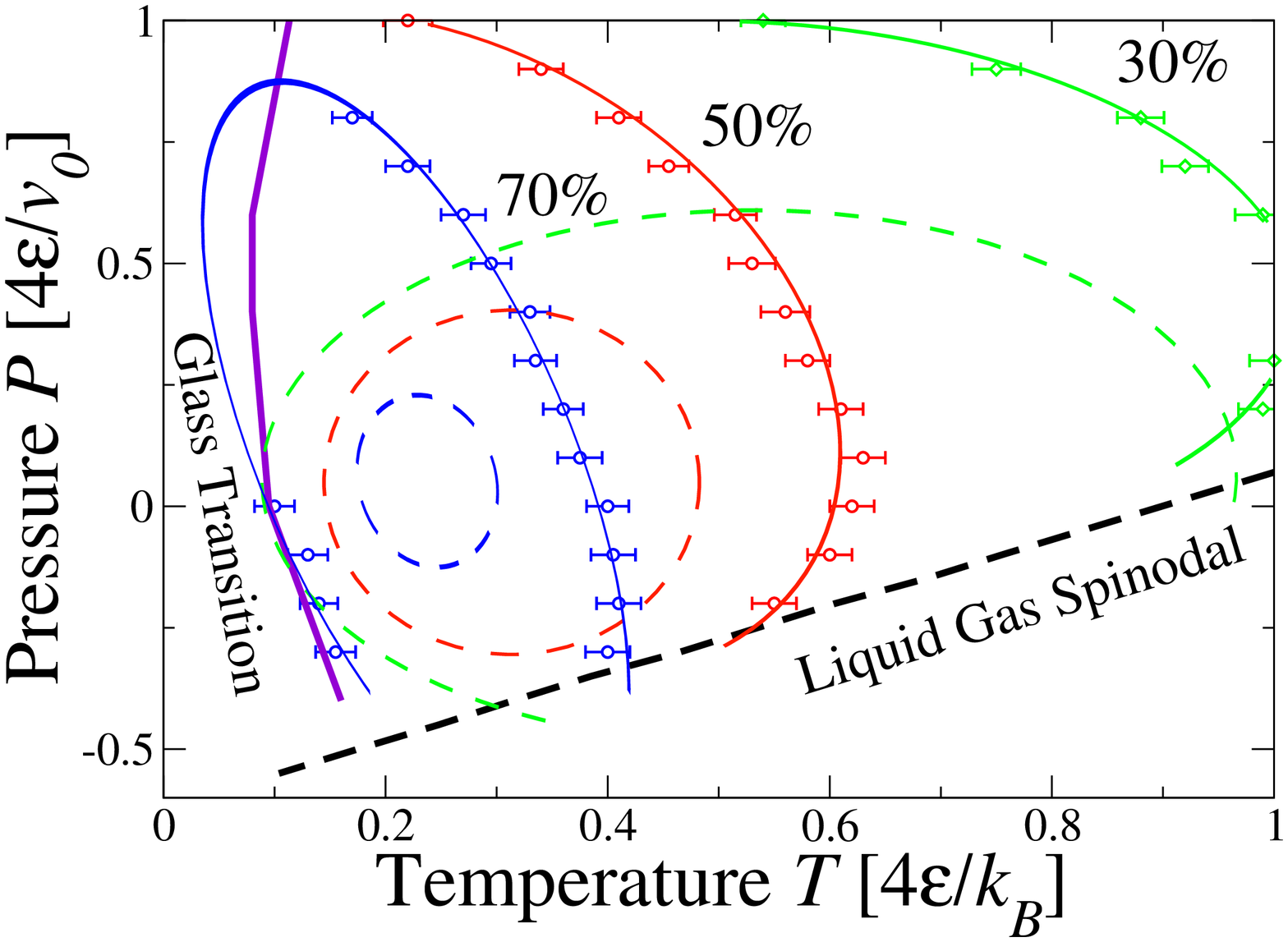}
\end{center}
\begin{center}
\leftskip 0.4cm {\bf (b)}\\
\includegraphics[scale=0.39,bb=0 0 668 450,clip=true]{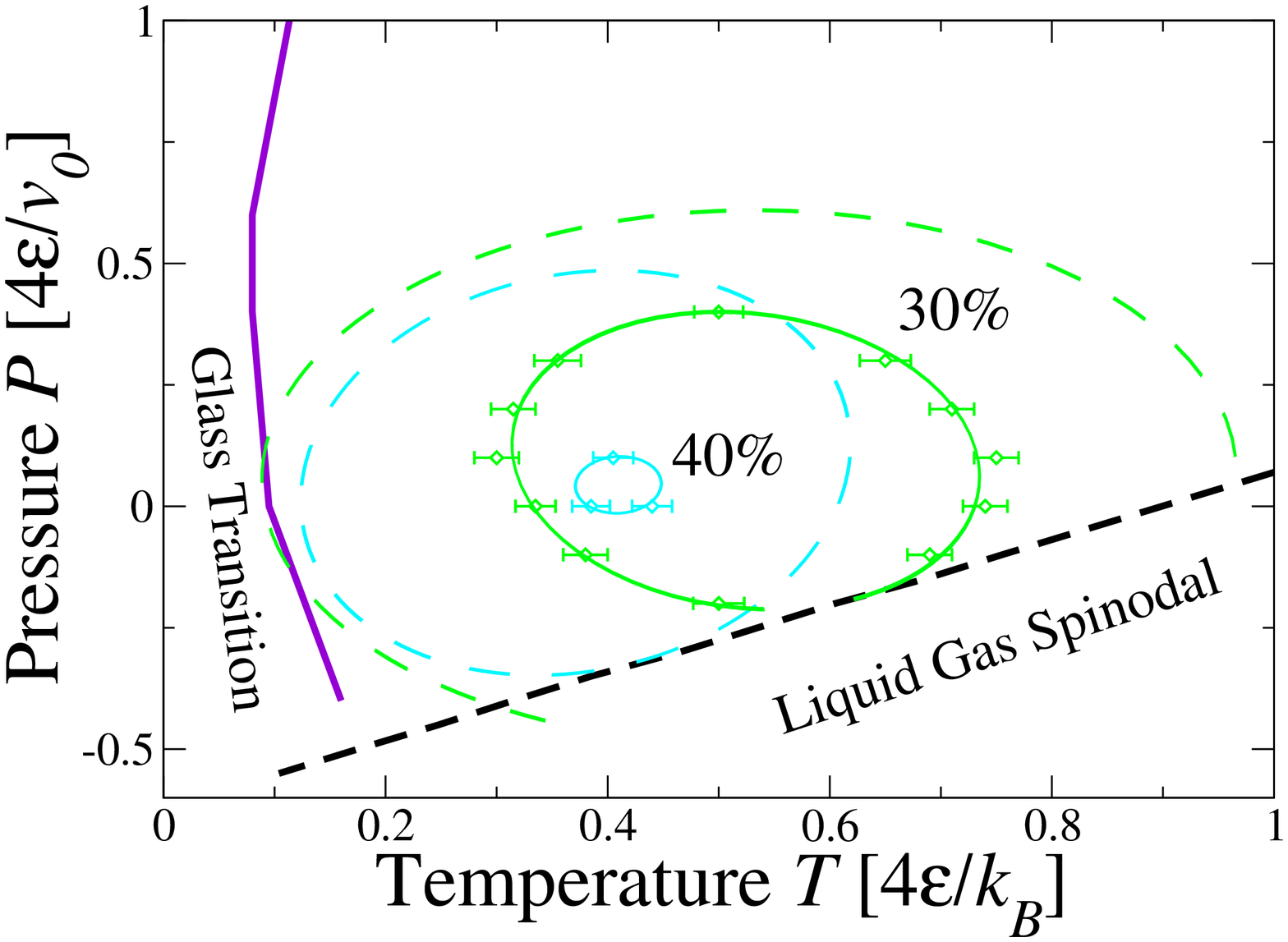}
\end{center}
\caption{Effect on the SR of changing the water-water HB directional component
  $J_\Phi$ at the $\Phi$-interface. In both panels symbols with
  continuous lines delimit the regions with 30\% (green),  40\% (turquoise),  50\% (red) and
  70\% (blue) of the protein folded.
  Dashed lines (with the same color code as for continuous lines) are
  for the reference system in Fig.\ref{SR0}  (Table \ref{Tab}) with
  $J_\Phi/4\epsilon=0.55$.
  All lines are guide for eyes.
  (a) For $J_\Phi/4\epsilon=0.20$, smaller than the reference value,
  the SR expands to lower $T$ and $P$ and  to higher $T$ and $P$.
  (b) For $J_\Phi/4\epsilon=0.75$, greater than the reference value,
  the SR shrinks.}
\label{SR_tune1}
\end{figure}

\subsubsection{Varying the water-water HB directional component  $J_\Phi$ at the $\Phi$-interface.}

Changing the (covalent) strength $J_\Phi$ of the interfacial HB
has a drastic effects on the SR. As discussed above, having
$J_\Phi/J>1$, as in the reference case, drives the cold unfolding as a consequence of the larger
gain of HB energy near the $\Phi$-interface. Instead, by setting  $J_\Phi/J<1$
(Fig.\ref{SR_tune1}a) the folded protein becomes more stable 
at low $T$ then in the reference case, because there is
a larger energy gain in forming as many bulk HB as possible, i.e., in
reducing the number of those near 
$\Phi$-residues. Hence, there is a larger free-energy gain in reducing
the exposed $\Phi$-interface with respect to the reference case.

As a matter of fact, with our choice
$J_\Phi/4\epsilon=0.20$, we find cold denaturation only for $P<0$.
This is a consequence of the fact that the
free energy has a term with $N_{\rm  HB}^{(\Phi)}$ multiplying 
$(-J_\Phi+Pv_{\rm  HB}-P^2v_{\rm  HB}v_0/4\epsilon)$, hence for $P<0$
the free energy decreases if $N_{\rm  HB}^{(\Phi)}$ increases, even
for a vanishing $J_\Phi$. The negative slope of the cold denaturation
line at $P<0$ (Fig.\ref{SR_tune1}a for 70\% curve) is because the
larger $|P|$, the larger is the  term proportional to
$N_{\rm  HB}^{(\Phi)}$ in the free-energy balance.

Reducing $J_\Phi$ makes the folded protein more stable also at
high $T$, because the entropy term
overcomes the energy term at  $T$ lower than in the reference
case. A similar observation holds also at high $P$, because a reduced
$J_\Phi$ implies a decrease in $N_{\rm  HB}^{(\Phi)}$, hence a
decrease in enthalpy gain associated to the exposure of the
$\Phi$-interface.

On the other hand, the larger $|P|$, the more negative is
the quadratic $P$-dependent coefficient that, as mentioned above,
multiplies $N_{\rm  HB}^{(\Phi)}$ in the free energy, and the larger
is the free-energy gain in exposing the $\Phi$-interface at high $T$. 
Hence, the hot-denaturation curve in the $P$-$T$ plane has a negative slope
for $P>0$ and a positive slope for $P<0$. As a consequence, the
ellipsis describing the SR (Fig.\ref{SR_tune1}a for 50\% curve)
becomes more elongated than in the 
reference case with a negatively-sloped major axis and an eccentricity
that grows toward 1.

On the contrary, for
 increasing  $J_\Phi$ the SR is lost, due to the energetic gain
 associated to 
  wetting the entire $\Phi$-interface of the protein
  (Fig. \ref{SR_tune1}b). The $P$-dependence of the  
  contour lines is the same as discussed for the case with
  $J_\Phi/J<1$, hence they  keep the shape but shrink.

\begin{figure}
\begin{center}
 \leftskip 0.4cm {\bf (a)}\\ 
\includegraphics[scale=0.39,bb=0 0 668 450,clip=true]{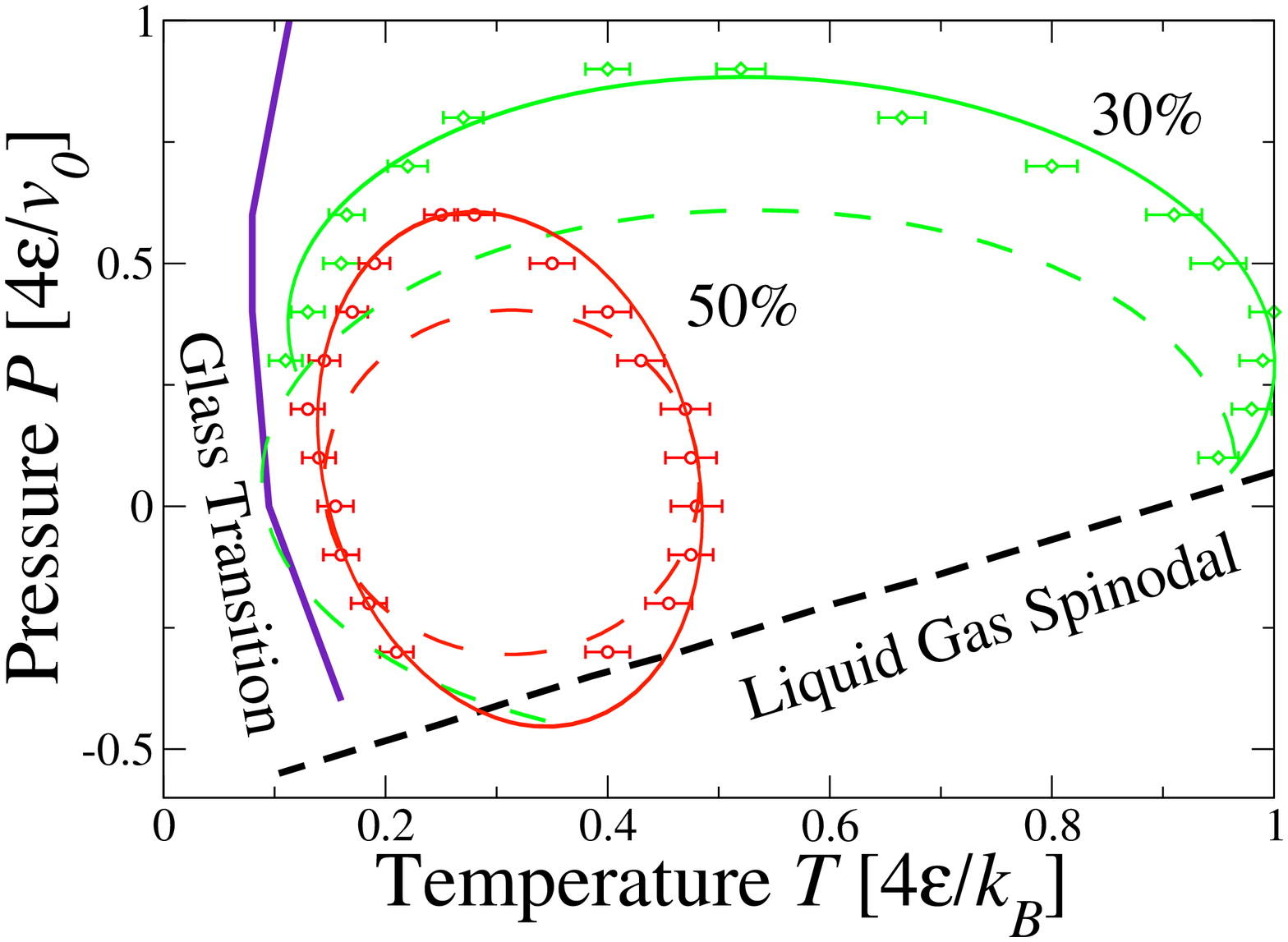}
\end{center}
\begin{center}
 \leftskip 0.4cm {\bf (b)}\\ 
\includegraphics[scale=0.39,bb=0 0 668 450,clip=true]{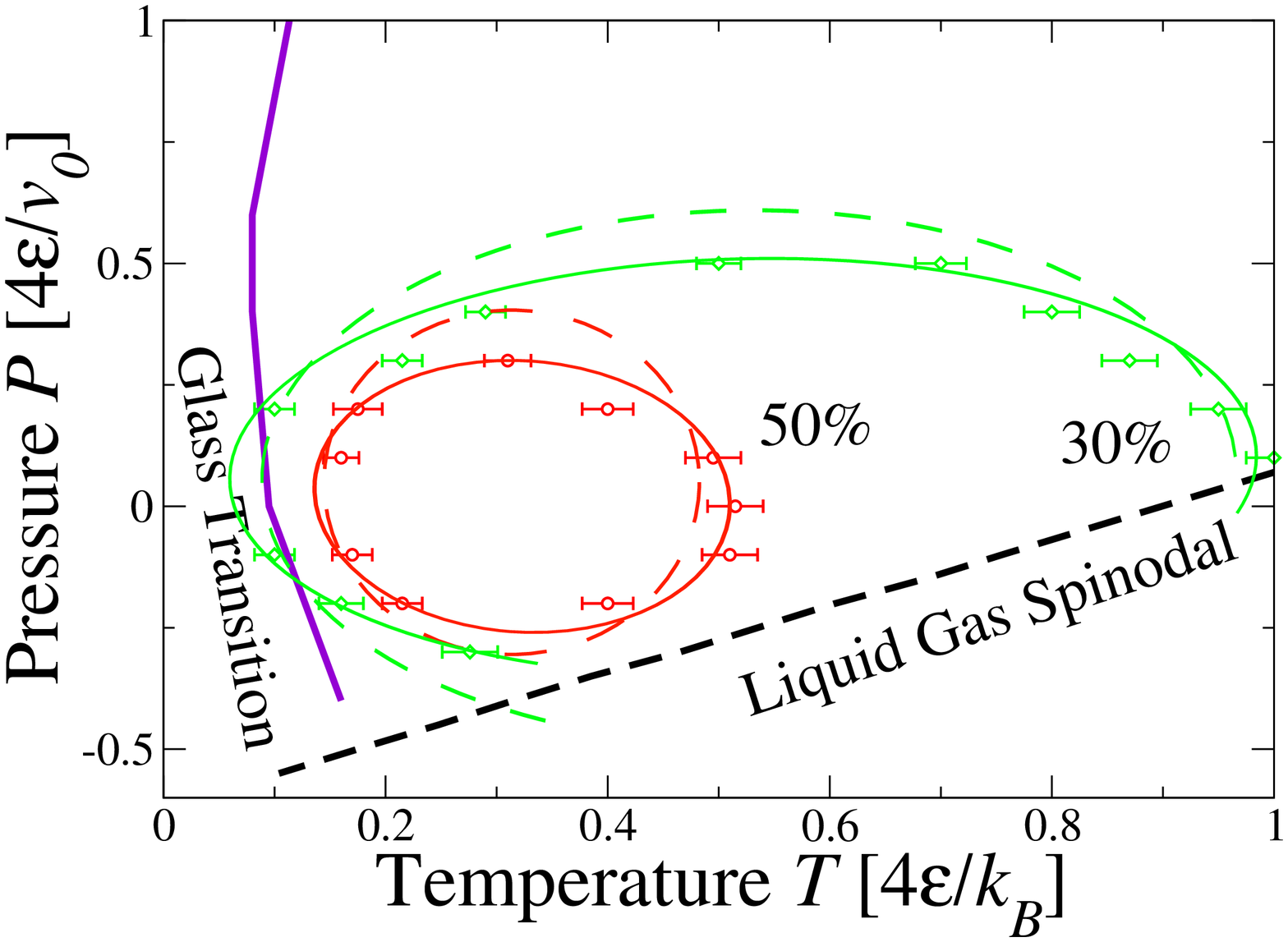}
\end{center}
\caption{Effect on the SR of changing the water compressibility factor
  $k_1$ at the  $\Phi$-interface. Symbols and lines are as in
  Fig.\ref{SR_tune1} and the reference system has 
  $k_1 4\epsilon/v_0=1$.
  (a) For $k_1 4\epsilon/v_0=0.5$,  smaller than the reference value,
  the SR expands to a wider range of $P$ and the main ellipsis axis
  acquires a negative slope in the $P$-$T$ plane.
  (b) For $k_1 4\epsilon/v_0 = 1.5$, greater than the reference value,
  the SR contracts in $P$ and the main ellipsis axis becomes almost
  perpendicular to the $P$-axis. In both panels the effects of the change on
  the $T$-range of stability are minor.}
\label{SR_tune2}
\end{figure}

\begin{table*}[]
  \centering
  \begin{tabular}{| c | c | c | c | c | c | c | c |}
    \hline
    $v_{\rm  HB}^{\rm (b)}/v_0$    &  $J/4\epsilon$    &  $J_\sigma/4\epsilon$
    & $v_{\rm HB,0}^{(\Phi)}/v_0$    & $J_{\Phi}/4\epsilon$     &
    $J_{\sigma,\Phi}/4\epsilon$
    & $k_1(4\epsilon)/v_0$    & $k_2=k_3$\\
    \hline
    \hline
    0.5    & 0.3    &    0.05
    & 0.5    & 0.55    & 0.05
    & 1    & 0\\
    \hline
  \end{tabular}
  \caption{\label{Tab} Parameters for the reference system of
    the hydrophobic protein model (Fig. \ref{SR0}) with which we compare the results
    after varying the constants at the $\Phi$-interface one by one.  
    We fix $v_0\equiv r_0^3=24.4$ \AA$^3$~ and    $\epsilon =5.8$ kJ/mol.}
  \end{table*}

\subsubsection{Varying the water compressibility factor $k_1$ at the
  $\Phi$-interface.}

Decreasing the water compressibility factor $k_1$ leads to a
stretching of the SR along the $P$ direction and a rotation of the
ellipse axes in a such a way that the main axis increases its negative
slope in the $P$-$T$ plane 
(Fig. \ref{SR_tune2}a). On the other hand, 
increasing $k_1$ results in a contraction of the SR along $P$ with a
rotation of the main axis toward a zero slope in the $P$-$T$ plane (Fig. \ref{SR_tune2}b).  

These effects can be understood observing that the free energy of the
system has a term $-k_1P^2N_{\rm  HB}^{(\Phi)}$. This term is associated to the fact
that there is a larger water compressibility at the $\Phi$-interface,
reducing the total free energy.  Therefore, by decreasing $k_1$ the
destabilizing effect of the increased water-compressibility is reduced
and the protein gains stability in $P$ at constant $T$, while the
opposite effect is achieved by increasing $k_1$. The observations
about the slope of the contour lines discussed in the previous
subsection apply also in this case explaining the rotation of the
ellipsis axes.

\subsubsection{Varying the HB volume-increase $v_{\rm HB, 0}^{(\Phi)}$
  at the  $\Phi$-interface and $P=0$.}

\begin{figure}
\begin{center}
 \leftskip 0.4cm {\bf (a)}\\ 
\includegraphics[scale=0.39,bb=0 0 668 450,clip=true]{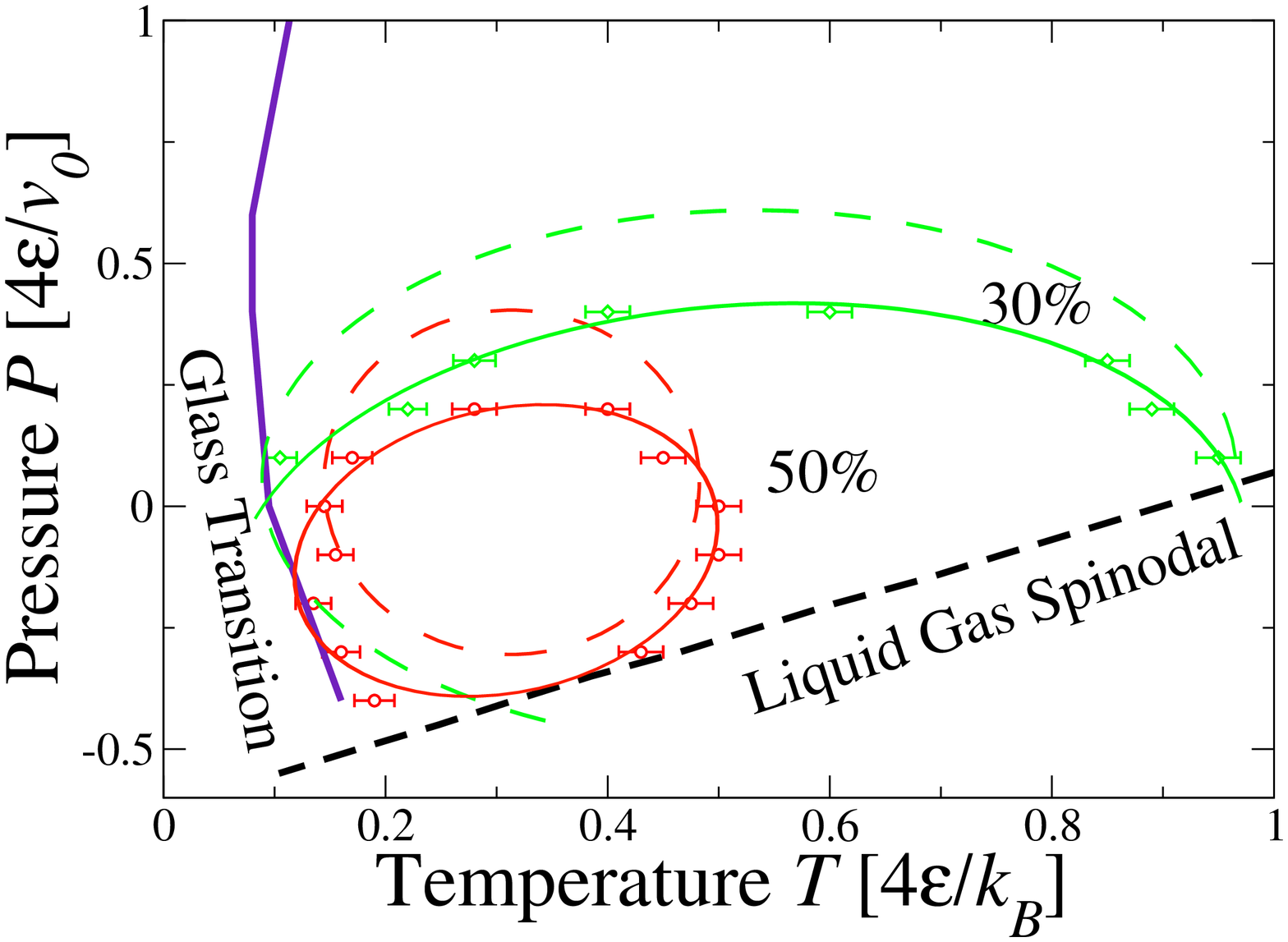}
\end{center}
\begin{center}
\leftskip 0.4cm {\bf (b)}\\ 
\includegraphics[scale=0.39,bb=0 0 668 450,clip=true]{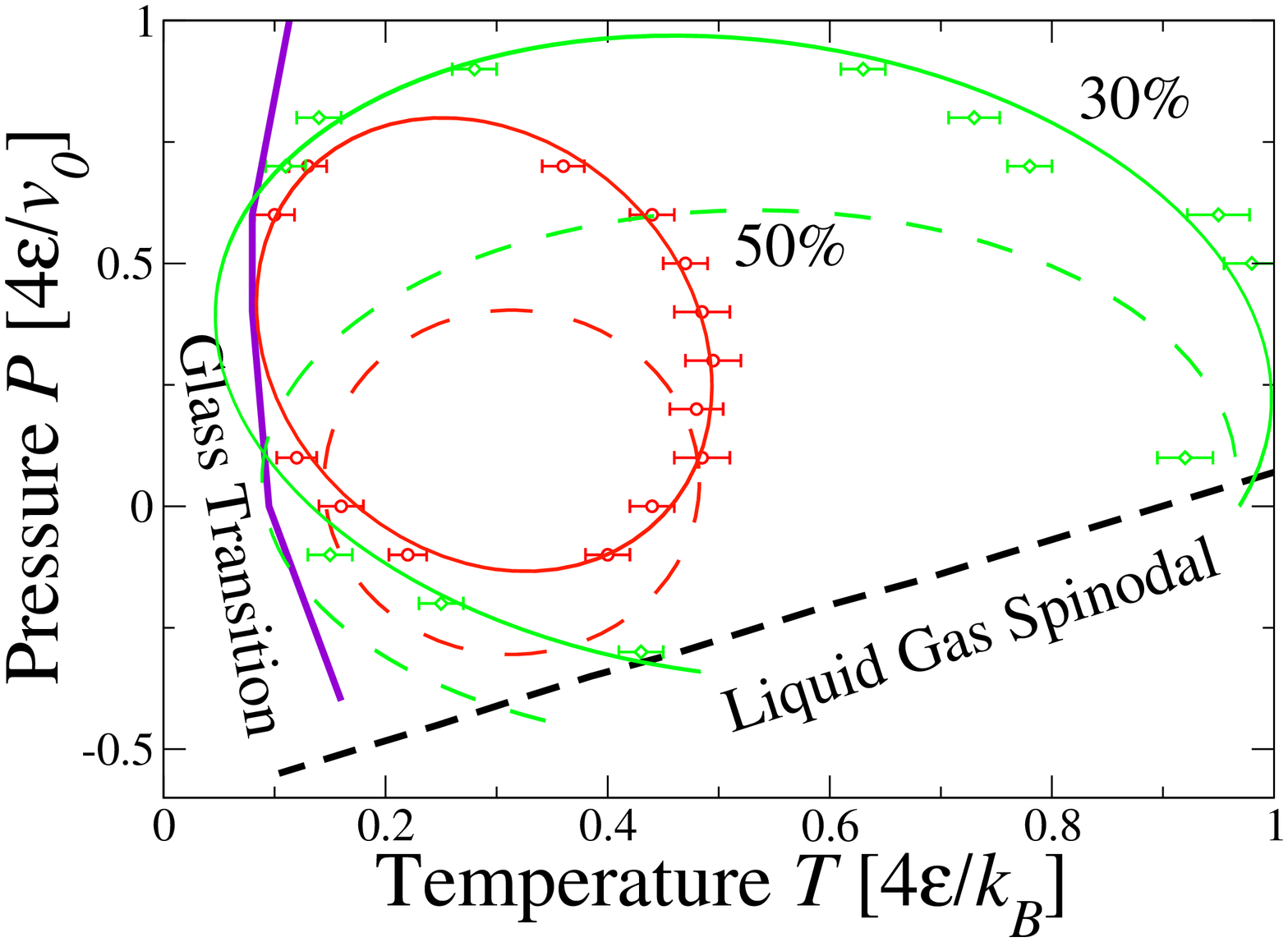}
\end{center}
\caption{Effect on the SR of changing the HB volume-increase $v_{\rm
    HB,0}^{(\Phi)}$  at the  $\Phi$-interface and $P=0$.
    Symbols and lines are as in  Fig.\ref{SR_tune1} and the reference
    system has   $v_{\rm   HB,0}^{(\Phi)}/v_0=0.5$.
  (a) For $v_{\rm  HB,0}^{(\Phi)}/v_0=0.1$,  smaller than the reference value,
  the SR moves toward lower $P$ and its main ellipsis axis rotates
  toward a positive slope in $P$-$T$ plane.
  (b) For $v_{\rm HB,0}^{(\Phi)}/v_0=1$,    greater than the reference value,
    the SR moves toward higher $P$ rotates
  toward a negative slope in $P$-$T$ plane.
       In both panels the effects of the change on
  the $T$-range of stability are minor.}
\label{SR_tune3*}
\end{figure}

A decrease  of $v_{\rm HB,0 }^{(\Phi)}/v_0$, respect to the reference
case, moves the SR at lower $P$, while an increase moves the SR at
higher $P$ (Fig. \ref{SR_tune3*}). This effect can be understood
observing that the free energy of the system has a term
$Pv_{\rm HB, 0}^{(\Phi)}N_{\rm  HB}^{(\Phi)}$ that, at each $P$, implies a
decreasing enthalpy cost for decreasing $v_{\rm HB, 0}^{(\Phi)}$ if
$N_{\rm  HB}^{(\Phi)}$ is kept constant. Hence, this term
favors  the unfolding at high $P$ when $v_{\rm HB, 0}^{(\Phi)}$ is small,
decreasing the stability of the native state upon pressurization
(Fig. \ref{SR_tune3*}a).  
The opposite occurs for increasing $v_{\rm HB, 0}^{(\Phi)}$
(Fig. \ref{SR_tune3*}b).  

We also find that the slope of the main ellipsis axis changes from
positive, for small $v_{\rm HB, 0}^{(\Phi)}$, to negative, for large
$v_{\rm HB, 0}^{(\Phi)}$. This is a consequence of the inversion of
the contribution of the free-energy term $Pv_{\rm HB, 0}^{(\Phi)}N_{\rm  HB}^{(\Phi)}$
when $P$ changes sign. Because a variation of $v_{\rm HB, 0}^{(\Phi)}$ changes
where the SR crosses  the $P=0$ axis, the stability contour-line
changes shape as a consequence, resulting in an effective  rotation of its
elliptic main axis: 
the main axis is positive when the majority of the SR is at $P<0$
(Fig. \ref{SR_tune3*}a) and is negative otherwise (Fig. \ref{SR_tune3*}b).  

\subsubsection{Adding the quadratic $P$-dependence of $v_{\rm HB}^{(\Phi)}$
  at the  $\Phi$-interface.}

So far we have shown the SRs for the model with $v_{\rm HB}^{(\Phi)}$
linearly-dependent on $P$. 
This truncation of  Eq. (\ref{vsurf}) implies that  the model 
for $P<1/k_1\equiv P_L$ describes a system where  water-water HBs at the
$\Phi$-interface decrease  the local density, as expected,
while for larger $P$ they do the opposite.
Thanks to our specific choice of parameters for the reference system,
our truncation
does not affects the results because for $P>P_L$ the 
HB probability, both in bulk and at the $\Phi$-interface, is
vanishing.

However, to check how qualitatively robust are our results against
this truncation of  Eq. (\ref{vsurf}),
we consider
also the case with the quadratic $P$-dependence of $v_{\rm
  HB}^{(\Phi)}$, i.e., 
\begin{equation}
 v_{\rm HB}^{(\Phi)}/v_{\rm HB,0}^{(\Phi)}\equiv 1- k_1P - k_2P^2,
 \label{vsurf2}
\end{equation} 
where $k_2>0$ is a new parameter with units of $k_1/P$.
With this new approximation of Eq. (\ref{vsurf}) results
$P_L\equiv (2/x) (\sqrt{1+x} -1)$, with $x\equiv 4k_2/k_1$. Therefore,
$P_L$  decreases for increasing $x$.

\begin{figure}
 \centering
\begin{center}
\includegraphics[scale=0.4, bb=0 0 668 450, clip=true]{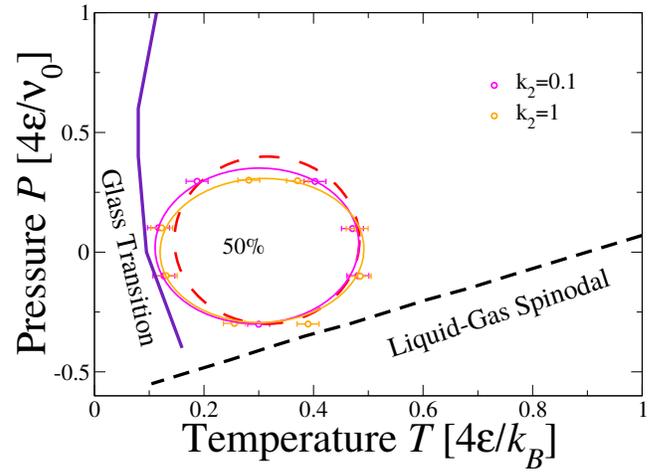}
\end{center}
\caption{Effect on the SR of adding the quadratic $P$-dependence of $v_{\rm HB}^{(\Phi)}$
  at the  $\Phi$-interface.      Symbols and lines are as in
  Fig.\ref{SR_tune1} and the reference system has
 $k_1=v_0^2/4\epsilon$ and $k_2=0$.
  For    $k_2(4\epsilon)^2/v_0^3 = 0.1$ (purple circles and line)
  with corresponding $P_Lv_0\simeq 0.98$,
   $k_2(4\epsilon)^2/v_0^3 = 0.5$ (not shown) with $P_Lv_0\simeq 0.90$
  and
  $k_2(4\epsilon)^2/v_0^3 = 1$ (orange circles and line)
  with $P_Lv_0\simeq 0.83$, the SR shrinks at high $P$ as $P_L$ decreases.}
\label{quadratic}
\end{figure}

We fix $k_1$ to the reference value, and vary $k_2$
(Fig. \ref{quadratic}). We find that for increasing $k_2$,
 the SR is progressively compressed on the high-$P$
 side, with minor effects on the SR $T$-range.
 Adding a cubic term in
Eq. (\ref{vsurf}) affects the SR in a similar way (data not shown).  
The rational for this behaviour lies in the enhanced enthalpic gain
upon exposing the $\Phi$-residue to the solvent since
$v_{\rm  HB}^{(\Phi)}$ decreases faster upon approaching $P_L$ that,
in turn, decreases for increasing $k_2$.

\begin{figure}
 \centering
\begin{center}
\includegraphics[scale=0.4,bb=0 0 668 450,clip=true]{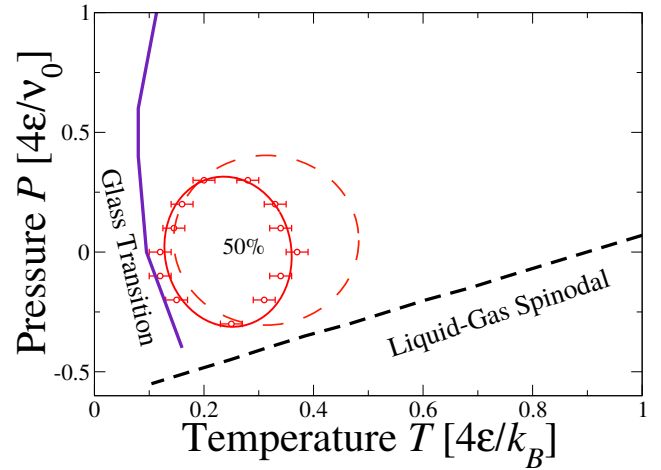}
\end{center}
\caption{Effect on the SR of adding  an attractive interaction $\epsilon_{\rm w,\Phi}$ between
  water and $\Phi$-residues.
  Symbols and lines are as in
  Fig.\ref{SR_tune1} and the reference system has
  $\epsilon_{\rm  w,\Phi}=0$.
  For $\epsilon_{\rm w,\Phi}/4\epsilon=0.05$, the SR moves toward
  lower $T$ and shrinks in $P$.} 
\label{referee_fig}
\end{figure}

\subsubsection{Adding an attractive interaction $\epsilon_{\rm w,\Phi}$ between
  water and $\Phi$-residues.}

Here, we check how a non-zero
water--hydrophobic residue interaction,
$\epsilon_{\rm w,\Phi}>0$, would affect the SR of the hydrophobic
homopolymer.
Indeed, despite the common misunderstanding of 
``water-phobia'' due to the oversimplified terminology,
it is 
well known that a hydrophobic interface
attracts water, but with an
interaction that is smaller than a hydrophilic surface.

We find that by setting $\epsilon_{\rm w,\Phi}/4\epsilon=0.05$,
smaller than bulk water-water attraction,
the SR is reduced in $P$ and lightly shifted toward lower $T$ (Fig. \ref{referee_fig}). 
In fact, an attractive
water--$\Phi$ interaction enhances the propensity of the polymer to
expose the $\Phi$ residues to the solvent, resulting in a global
reduction  of the SR and destabilizing the folded protein.  

\subsubsection{Enhancing the cooperative  interaction $J_{\sigma,\Phi}$   at the $\Phi$-interface.} 

\begin{figure}
 \centering
\begin{center}
  \includegraphics[scale=0.4, bb=0 0 668 445, clip=true]{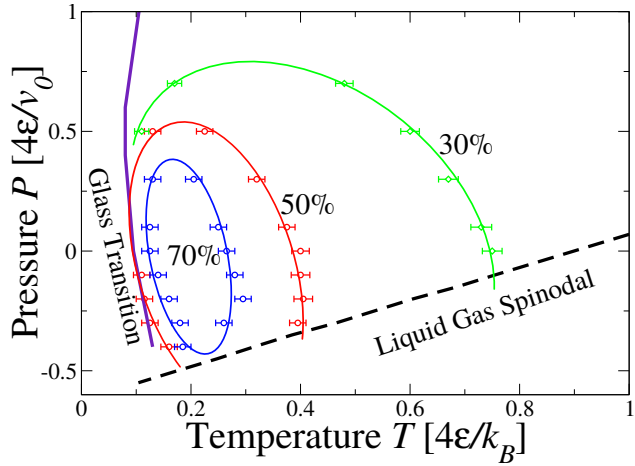}
\end{center}
\caption{Effect on the SR of enhancing the cooperative  interaction $J_{\sigma,\Phi}$   at the $\Phi$-interface.     Symbols and lines
  are as in Fig.\ref{SR_tune1}.
      Here we adopted $J_{\sigma,\Phi}/4\epsilon = 0.1$, twice the
      value of $J_\sigma$ for bulk water molecules, while we fix 
       $J_{\Phi}/4\epsilon=J/4\epsilon=0.3$. The
      parameters $k_1$ and  $v_{\rm  HB,0}^{(\Phi)}$ are as in
      Fig. \ref{SR_tune2}a.  
}   
\label{strong_coop}
\end{figure}

Lastly, in the contest of the hydrophobic protein model, we consider a different scenario. 
As discussed in the model description, the enthalpic
gain upon cold denaturation would be
consistent  also with the assumption
$J_{\sigma,\Phi} > J_\sigma$ associated to a larger cooperativity of
the HBs at the $\Phi$-interface. 
Hence, to analyze this scenario, we compute the SR considering the
directional component of the HB unaffected by the $\Phi$-interface
$J_{\Phi}=J$, while assuming an enhanced HB cooperativity at the
$\Phi$-interface $J_{\sigma,\Phi} > J_\sigma$.
Note that the increase of $J_{\sigma,\Phi}$ promotes the number
of cooperative HBs at the  $\Phi$-interface only once they are formed
as isolated HBs ($J_{\sigma,\Phi}<J_{\Phi}$).
Our finding (Fig. \ref{strong_coop}) are consistent with a close SR, presenting cold- and pressure-denaturation. 
 

Although not discussed here, we expect that varying the parameters $k_1$ and $v_{\rm  HB,0}^{(\Phi)}$, with the current choice of $J_{\sigma,\Phi} > J_\sigma$ and $J_\Phi = J$, would affect the SR similarly to the cases discussed in previous subsections.

\begin{figure}
\begin{center}
  \includegraphics[scale=0.4, bb=0 0 668 445, clip=true]{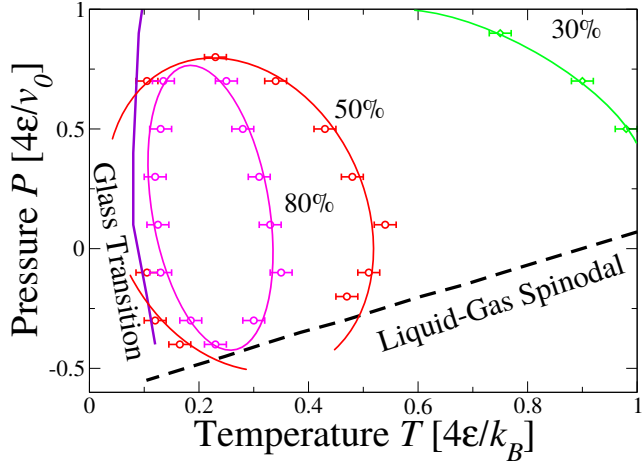}
\end{center}
 \caption{The SR for the polar protein model. We set the  parameters as in
   Table \ref{Tab2}
   with all the other parameters as in
   Table \ref{Tab}.
 Symbols with continuous lines delimit the regions
with 30\% (green), 50\% (red) and 80\% (magenta) of the
protein folded. The other lines are as in Fig.~\ref{SR0}. All lines are guides for eyes.}
\label{improved_model}
\end{figure}

\subsection{Results for the polar protein model.}

Next we summarize the results for the polar protein model.
As shown in Ref. \cite{BiancoPRL2015}, also in this case the SR
recover a close elliptic--like SR in the $T$--$P$ plane
(Fig. \ref{improved_model}). In particular, despite we reduce the value of
$J_{\Phi}/4\epsilon$ with respect to  the hydrophobic
protein model in Table \ref{Tab}, the additional residue-residue
interaction $\epsilon_{\rm  rr}$ and water--$\zeta$-residue
interaction $\epsilon_{\rm w,\zeta}$ stabilize the folded state to
higher $P$ and $T$, as can been seen by comparing
Fig.~\ref{improved_model} with Fig.~\ref{SR0}.

\begin{figure}
\begin{center}
 \leftskip 0.4cm {\bf (a)}\\ 
 \includegraphics[scale=0.39, bb=0 0 668 445, clip=true]{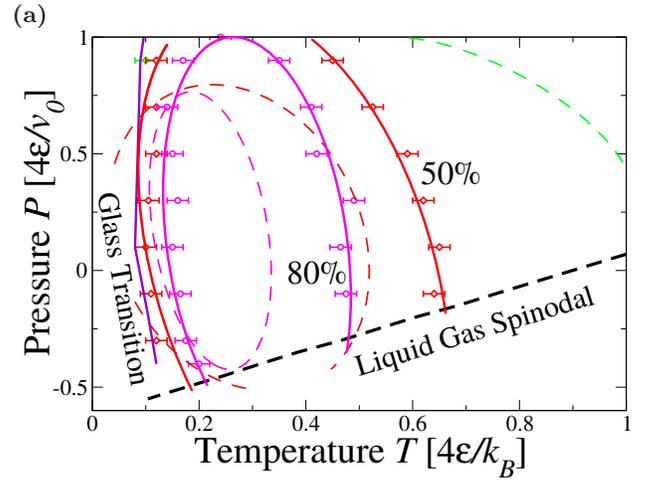}
\end{center}
\begin{center}
  \leftskip 0.4cm {\bf (b)}\\ 
  \includegraphics[scale=0.39, bb=0 0 668 445, clip=true]{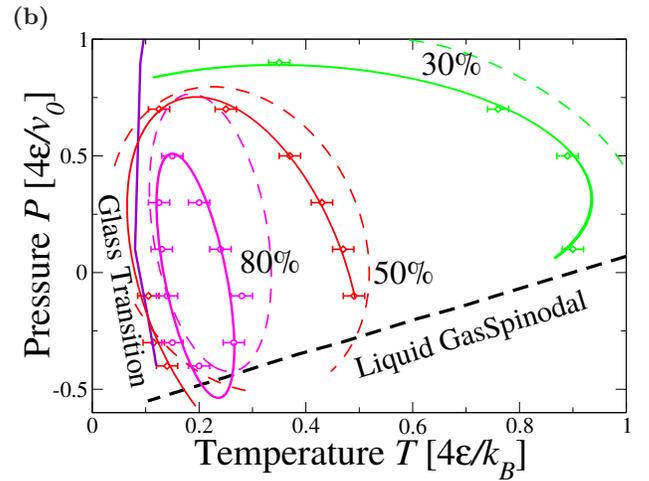}
\end{center}
 \caption{Effect on the SR of the polar protein of varying 
   the residue--residue interaction $\epsilon_{\rm rr}$.
   In both panels dashed lines  are for the reference system
   in Fig. \ref{improved_model} (Table \ref{Tab2}) with
   $\epsilon_{\rm rr}/4\epsilon=0.2$,
   continuous (with
   the same color code as for dashed lines) are for the systems with a
   modified  $\epsilon_{\rm rr}$.
   (a) For $\epsilon_{\rm rr}/4\epsilon=0.5$, greater than the
   reference value, the SR expands  in $P$ and $T$.
  (b) For $\epsilon_{\rm rr}/4\epsilon=0.05$, smaller than the
   reference value,  the SR reduces in $P$ and $T$.  }
 \label{improved_model_mm}
\end{figure}

\subsubsection{Varying the residue-residue interaction $\epsilon_{\rm rr}$.}

To test how the residue-residue interaction $\epsilon_{\rm rr}$ is relevant for stabilizing
the folded protein, we change its value. We find that  an increase of 
$\epsilon_{\rm rr}$ results in a broadening of the SR in 
$T$ and $P$ (Fig~\ref{improved_model_mm})a.
We find the opposite effect  if we reduce $\epsilon_{\rm rr}$ (Fig
\ref{improved_model_mm}b). 
These results are consistent with
our understanding that  the native state is stabilized by stronger residue-residue
interactions.

\subsubsection{Varying  $J_{\Phi}/4\epsilon$
    and $v_{\rm  HB,0}^{(\Phi)}$  at the  $\Phi$-interface.}

Next, we evaluate the effects of changing the water-water
$J_{\Phi}/4\epsilon$ interaction and the HB volume increase constant
$v_{\rm  HB,0}^{(\Phi)}$ at the  $\Phi$-interface  for the polar
protein model. We find that these changes affect the SR in a fashion
similar to those discussed for the hydrophobic protein model (not shown).

\begin{table}[]
  \centering
  \begin{tabular}{| r | c | c | c | c | c | c |}
    \hline
    Protein
    &   $\epsilon_{\rm rr}/4\epsilon$
    &  $\epsilon_{\rm w,\Phi}$
    &  $\epsilon_{\rm w,\zeta}/4\epsilon$
    & $J_{\Phi}/4\epsilon$
    & $v_{\rm  HB}^{(\zeta)}$
    & $J_{\zeta}/4\epsilon$\\
    \hline
    Polar
    & 0.2
    & 0
    &   0.35
    & 0.5
    & 0
    & 0.4\\
    \hline
  \end{tabular}
  \caption{\label{Tab2} Additional parameters for the reference    systems     of
    the polar protein model (Fig. \ref{improved_model}) with respect to those of the hydrophobic
    protein model in Table \ref{Tab}. We also reduce the value of
    $J_{\Phi}/4\epsilon$ with respect to Table \ref{Tab}. }
  \end{table}

\section{Perspective on the protein design}

As we mentioned in the previous sections, the hydrated protein models
discussed here simplify the dependence of the stability against
unfolding on the protein sequence. In fact, in the homopolymer protein model, the
sequence is reduced to a single amino acid, hence we have the alphabet
$\mathscr{A} = 1$, while in the polar protein model the alphabet size
coincides, by construction, with the protein length $l$,
$\mathscr{A} = l$, because the interaction matrix has $(l^2 -l)/2$
different elements that depend on the native state configuration.

In a more realistic case we would deal with proteins composed, at most,
by 20 different amino acids, irrespective of the protein length.
The amino acids assemble in a linear chain, which defines the
protein sequence, in such a way that the protein is capable to fold
into a unique native structure. Usually, among the huge amount of
possible sequences, only few are good folders for a given native
structure, smoothing and funneling the free energy landscape in order
to lead the open protein conformation toward the native one.

Protein design strategies allow us
to identify good folding sequences for each native conformation.
Different
methodologies have been proposed and studied in the
past years \cite{Desjarlais1995,smith1995, harbury1998,
  DahiyatScience1997, Dahiyat1997, Kuhlman2003, noble2004,
  Borovinskiy2003, Berezovsky2007, Rothlisberger2008, Das2008,
  Jiang2008,  King2012, Boyken2016, Jacobs2016a} but water properties
are not explicitly accounted, apart from few cases
\cite{Desjarlais1995, DahiyatScience1997, Dahiyat1997, Kuhlman2003,
  Rothlisberger2008, Jiang2008} usually referred only to ambient
conditions. Despite the fact that the evolution has selected natural
protein sequences capable to fold and work in extreme thermodynamic
conditions (like the anti-freeze proteins or the thermophilic
proteins), all the design methods are not efficient in establishing
which are the key elements to predict artificial sequences stable in
thermodynamic conditions far from the ambient situation.

On this important aspect our model can give a relevant
insight. Indeed, following the works of Shakhnovich and 
Gutin \cite{Shakhnovich1993c, Shakhnovich1993a} on lattice proteins,
we can easily introduce an interaction matrix between the 20 amino
acids---like the   
Miyazawa Jernigan residue-residue interaction matrix $S$
\cite{Miyazawa1985}---and look for the protein sequences which
minimize the energy of the native structure. This scheme can be
improved to account for the water properties of the surrounding water,
since the protein interface affects the water-water hydrogen bonding
at least in the first hydration shell. In this way, we aspect to find
sequences with patterns depending on the $T$ and $P$ conditions of the
surrounding water. Our preliminary results show that the protein
sequences designed with our explicit-water model
strongly depend on the thermodynamic conditions of the
aqueous environment.

\section{Conclusions}

In this work we have presented a protein--water model to investigate
the effect of the energy and density fluctuations at the hydrophobic
interface ($\Phi$) of the protein. In particular, we have considered
two protein models. In the first we simplify the discussion assuming
that the protein is a hydrophobic homopolymer. In the second model we
consider a more realistic case, assuming that the protein has
a unique native state with a hydrophilic ($\zeta$) surface and a
hydrophobic core and that the hydrophilic residues polarize the
surrounding water molecules. 
In both cases, we model the hydrophobic effects considering that the
water--water hydrogen bond at the $\Phi$-interface
 are stronger with
respect to the bulk, and that the corresponding density fluctuations
are reduced upon pressurization.

Our model qualitatively reproduces the melting, the
cold-- and the pressure--denaturation experimentally observed in
proteins. The stability region, i.e. the $T$--$P$ region where the
protein attains its native state, has an elliptic--like shape in the
$T$--$P$ plane, as predicted by the theory \cite{hawley}.

We discuss
in detail how each interaction affects the stability region, showing
that our findings are robust with respect to model parameters changes.
Aiming at summarize our findings, although
the parameter variations results in a non trivial modification of the
protein stability region, we observe
that the strength of the interfacial water-water HB compared to the
bulk ones, mainly affect the $T$--stability range of proteins, while
the compressibility of the hydrophobic hydration shell mainly
regulates the $P$--stability range.
The scenario remain substantially
unvaried by changing the protein model from the oversimplified 
hydrophobic homopolymer to the 
polar protein model.
Our findings put water's density and energy fluctuations in a primary
role to mantain the stable protein structure and pave the way for a
water--dependent design of artificial proteins, with tunable
stability.

\subsection*{Acknowledgments}

V.B. acknowledges hospitality at Universitat de Barcelona during his
visits and the support from the
Austrian Science Fund (FWF) project  M 2150-N36.
V.B. and I.C.
acknowledge the support from the  Austrian Science Fund
(FWF) project P 26253-N27.
G.F. acknowledges the support from the
FIS2015-66879-C2-2-P (MINECO/FEDER) project.
V.B., I.C. and G.F.
acknowledge the support of the Erwin Schr\"odinger
International Institute for Mathematics and Physics (ESI). 




\end{document}